\documentclass[11pt,twoside]{article}

\usepackage{msf07}
\usepackage{graphicx}

\begin{document}
\title{The starburst cluster Westerlund 1 and its Galactic siblings -- Observation confronts Theory}   
\author{Wolfgang Brandner}   
\affil{Max-Planck-Institut f\"ur Astronomie}    

\begin{abstract} 
Because of their large number of stars spread over the entire stellar mass spectrum, starburst clusters are highly suitable to benchmark and calibrate star formation models and theories.
Among the handful of Galactic starburst clusters, Westerlund 1 with its estimated 150 O-stars, W-R stars, supergiants and hypergiants is the most massive young cluster identified to date in the Milky Way. While previous studies of Westerlund 1 focused largely on optical and X-ray observations of its evolved massive stellar population, we have analyzed near-infrared data, resulting in the first in depth study of the ``lower-mass'' main sequence and pre-main sequence cluster population, i.e., of stars in the mass range 0.4 to 30 solar masses. 
The derived properties of the cluster population allow us to test theoretical evolutionary tracks. By comparison of Westerlund 1's half-mass radius with younger starburst clusters like NGC 3603\,YC and Arches, and somewhat older massive clusters like RSGC1 and RSGC2, we find evidence for a rapid dissolution of Galactic starburst clusters, which has interesting implications for the long-term survival of starburst clusters, and the question to which extent Galactic starburst clusters might mimic proto-globular clusters. 

\end{abstract}


\section{Introduction}   
Starburst clusters with ages of a few million years represent unique astrophysical laboratories, as stars across the entire stellar mass range from the upper mass cut-off in the mass function down to the hydrogen burning limit (and possibly beyond), and with the same metallicity and age are present in a rather homogeneous environment.
As such, starburst clusters are the ideal places to study star formation and to test theories on stellar and cluster formation and evolution. Unlike interacting galaxies like the Antennae galaxies, where 100s of starburst clusters have been identified (Whitmore \& Schweizer 1995), the Milky Way houses only a handful of starburst clusters. Starburst clusters in the Antennae, however, are barely resolved, restricting us to study the integrated properties of 100,000s of stars. In the Milky Way, on the other hand, starburst clusters can be resolved into 1,000s to 10,000s of stars, and the properties of each star can be derived individually. 

\subsection{What defines a starburst cluster?}

What exactly is a ``starburst'' and what defines a ``starburst cluster''? 

In general, a starburst is a special place in space and time, where a spike in the star formation rate significantly above the ``average rate'' is observed. The Orion nebula cluster is often quoted as an example for a nearby starburst. In the following, we are concentrating on considerably more extreme star formation events. In our (admittedly somewhat loose) definition, a starburst cluster is one of the most extreme star formation environments found in present-day Milky Way with the following properties:\\

A starburst cluster
\begin{itemize}
\item[$\bullet$] contains at least several 10,000 stars
\item[$\bullet$] contains at least 10,000 M$_\odot$ in stellar mass
\item[$\bullet$] houses massive stars with initial masses of $\approx$120\,M$_\odot$ (corresponding to an MK-type of O2V)
\end{itemize}

\subsection{Galactic starburst cluster}

\begin{figure}[htb]
\centerline{
\includegraphics[width=10cm]{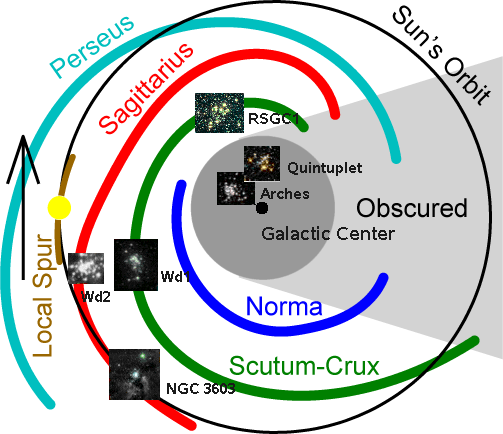}}
\caption{Location of the presently-known starburst clusters plotted on a map (courtesy of Wikipedia) of the Milky Way spiral arm structure.
The Sun's orbit is indicated by a black circle, and the present-day position of the Sun by a yellow dot. The small inserts show near-infrared observations of the individual starburst clusters (see main text for references). \label{wb_fig1}}
\end{figure}

Milky Way starburst clusters can be found in two different environments. The $\approx$2\,Myr old Arches cluster and the 3--6\,Myr old Quintuplet cluster are located near the center of the Milky Way with projected separations of less than 100\,pc from the Galactic Center. The 1\,Myr old NGC 3603 young cluster (YC), and the 3--5\,Myr old clusters Westerlund 1 (Brandner et al.\ 2008) and 2 (Ascenso et al.\ 2007) are located in spiral arms at distances of 5 to 8\,kpc from the Galactic center. Because of strong extinction and high stellar density in the Galactic plane, our census of Galactic starburst clusters is most likely incomplete, as is also high-lighted by the recent discovery of the two embedded red supergiant clusters RSGC1 and RSGC2 located in the Scutum-Crux spiral arm (Davies et al.\ 2007). 5 to 10\,Myr ago, RSGC1 and RSGC2 would most likely have qualified as starburst clusters according to our definition.

\subsection{Advantages of studying spatially resolved starburst cluster}

\begin{figure}[htb]
\centerline{
\includegraphics[width=8cm]{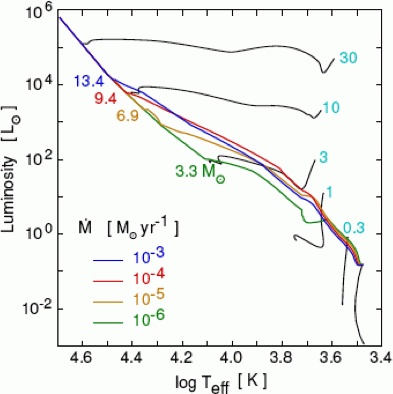}}
\caption{Pre-main sequence evolutionary tracks for stars with a range of masses. Without accretion, stars of a given mass follow the black tracks. In the presence of on-going accretion during pre-main sequence evolution, the tracks change quite significantly (courtesy of H.W.\ Yorke) \label{wb_fig2}}
\end{figure}

There are several advantages in studying local, and hence spatially resolved starburst clusters.
The large number of stars is crucial for a statistically sound determination of the mass function and dynamical properties of the clusters. 

Compared to less extreme star formation environments, starburst clusters initially house the most massive and luminous O-type main sequence stars. UV photons from these massive stars lead to rapid photo-evaporation of any remnant circumstellar material around the low-mass members of the cluster. This in turn brings two advantages. First, there is very little differential extinction and IR excess. While this is bad news for anyone looking for circumstellar disks and planets, it results in a well constrained colour-magnitude sequence for the cluster members. Secondly, the absence of circumstellar material means that non-accreting pre-main sequence tracks can be used to compare theory with observations. As, e.g., shown by Siess et al.\ (1997) or Zinnecker \& Yorke (2007),, the presence of on-going accretion alters pre-main sequence evolutionary tracks quite drastically (Figure \ref{wb_fig2}).

\section{Westerlund 1 - Testing evolutionary tracks}   

The following analysis is based on near-infrared observations of Westerlund 1. Seeing limited wide-field data obtained with the ESO NTT and SOFI cover an area of $\approx 5{\rm pc} \times 5{\rm pc}$ centered on Westerlund 1 (Brandner et al.\ 2008). This is supplemented by adaptive optics high-resolution imaging of the cluster center obtained with NACO at the ESO VLT.

\begin{figure}[htb]
\centerline{
\includegraphics[width=13cm]{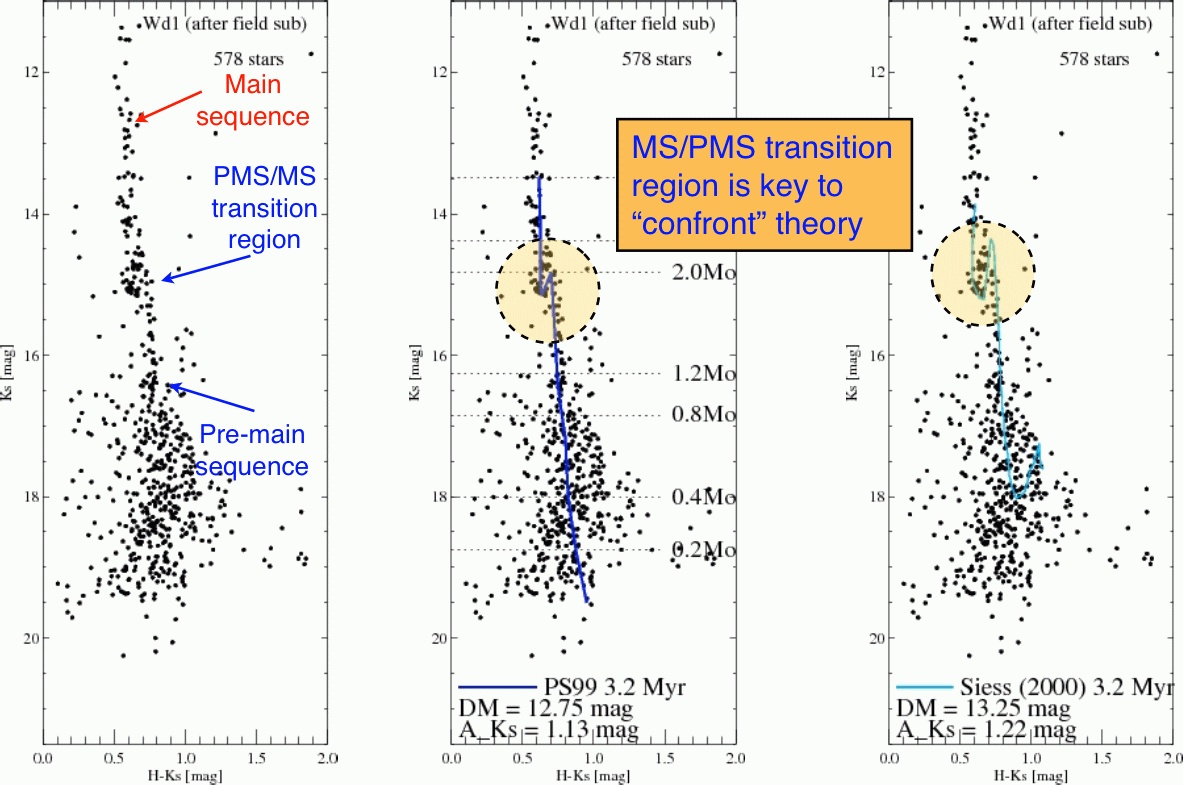}}
\caption{Near-infrared colour-magnitude diagramme of the central region of Westerlund 1 obtained with adaptive optics (NACO) at the ESO VLT. In the figure on the left, the pre-main sequence (PMS) and main sequence (MS) population as well as the PMS/MS transition region are identified. In the middle, the best fitting isochrone PS99 from Palla \& Stahler (1999) is overplotted. The PS99 isochrone provides both a good fit to the transition region and yields the same value for the foreground extinction as has been determined by a comparison of main-sequence stars with a Geneva isochrone. The figure on the right highlights that an isochrone by Siess et al.\ (2000) does not fit the transition region as well. The offset in infrared intrinsic colours for the lower mass MS stars when compared with Geneva isochrones indicates a potential problem in the transformation from the theoretical to the observational plane for the Siess tracks. \label{wb_fig3}}
\end{figure}

Figure \ref{wb_fig3} shows a Ks vs.\ H--Ks NACO colour magnitude diagramme for a $0.5{\rm pc} \times 0.5{\rm pc}$ central field located just to the east of the center of Westerlund 1. Unrelated foreground and field stars has been statistically subtracted based on the observations of the comparison field. The well defined cluster main-sequence, transition region and pre-main sequence are indicated. Theoretical evolutionary tracks and the derived isochrones differ in particular in their prediction of the transition region. In the present example, tracks by Palla \& Stahler (1999) give a better fit to the observations than tracks by Siess et al.\ (2000). Still, a finer mass-sampling of the PS99 tracks would be required for a more detailed calibration against the PMS/MS transition region.


\section{Starburst clusters going bust - or are they proto-globular clusters?}   

Given a total stellar mass of at least several 10,000 solar masses, starburst clusters must have formed out of giant molecular clouds. Once the most massive stars appear on the main sequence, they rapidly ionize and disperse the remaining gas. Simulations by a variety of groups (e.g.\ Hills 1980, Lada et al.\ 1984, Geyer \& Burkert 2001) indicate that in general a star formation efficiency (SFE) of at least 30\% is required for a stellar cluster to remain bound, though under special circumstances a SFE as low as 10\% might suffice for clusters to survive for 100\,Myr (Adams 2000, Baumgardt \& Kroupa 2007).

To answer the question if any of the local starburst clusters constitutes a proto-globular cluster, observation of the cluster kinematics are required. Thus far 1d velocity dispersions derived from radial velocity measurements of a handful of the brightest cluster members in Arches (Figer et al. 2002) and Westerlund 1 (Mengel \& Tacconi-Garman 2007) have been obtained, and - assuming virial equilibrium - employed to estimate an upper limit of the total mass in each of the two clusters.

\begin{figure}[htb]
\centerline{
\includegraphics[width=13cm]{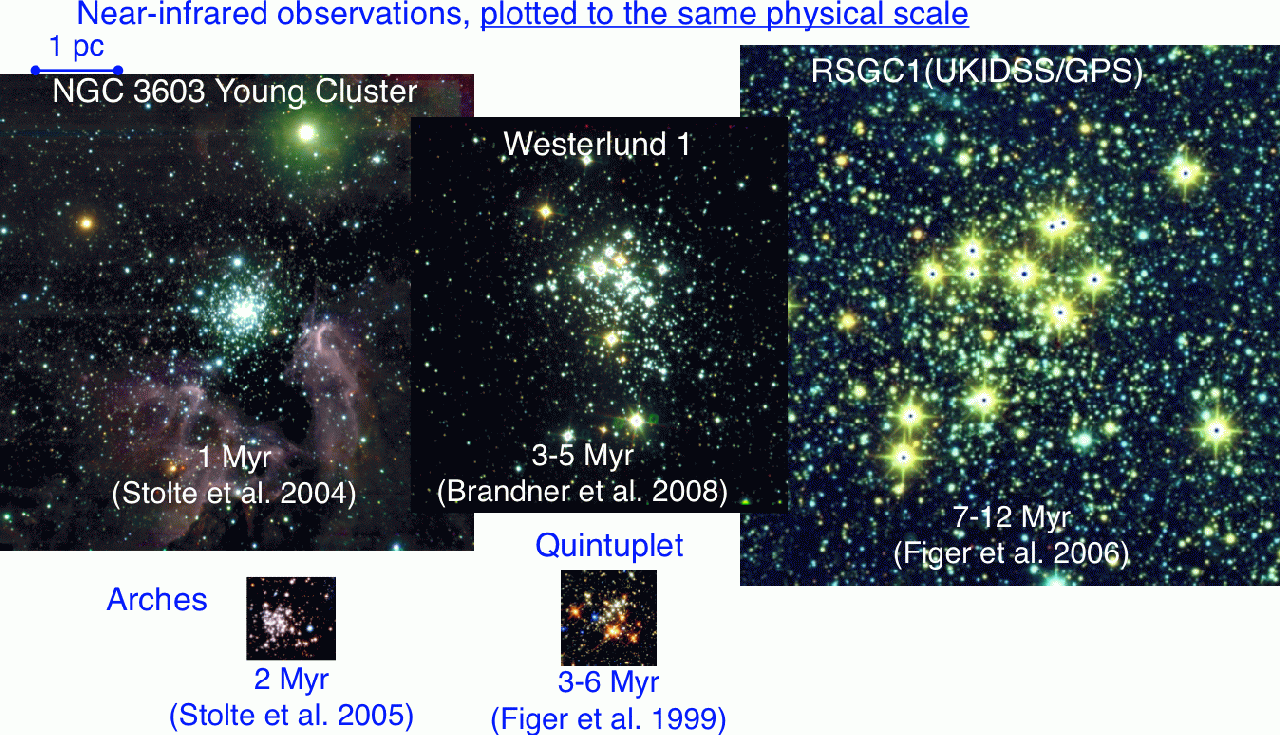}}
\caption{Picture gallery of near-infrared observations of Galactic starburst clusters all plotted to the same physical scale, ordered according to their age from left to right. Also shown is the recently identified red super giant cluster RSGC1 with an age of $\approx$10\,Myr. The apparent increase in cluster size as measured by the half-mass radius with increasing age is suggestive of a rapid dynamical evolution (and dissolution in the general Galactic field) of the starburst clusters.\label{wb_fig4}}
\end{figure}

Figure \ref{wb_fig4} shows a picture gallery of near-infrared observations of Galactic starburst clusters all plotted to the {\it same} physical scale, and ordered according to age from left to right. Only NGC 3603 YC and Arches, the two youngest clusters in the sample, exhibit compact cores with half-mass radii of less than 0.5\,pc, whereas the already slightly more evolved Westerlund 1 and Quintuplet clusters have half-mass radii of 1\,pc. The two recently discovered red super giant clusters (Figer et al.\ 2006; Davies et al.\ 2007) with ages of around 10\,Myr have still larger half-mass radii. 
For the spiral arm clusters, which are experiencing only weak tidal fields, this could be evidence that dynamical evolution is accelerated by the gas expulsion (see Baumgardt \& Kroupa 2007 for recent simulations). For the starburst clusters in the Galactic center region, strong tidal shear could result in rapid cluster dispersal (Kim et al.\ 2000; Portegies-Zwart et al.\ 2002).

Thus there are hints that the current generation of Milky Way starburst clusters is not long-lived, and hence is different from proto-globular clusters.

\section{Outlook}   

Recently, Stolte et al.\ (2008) compared multi-epoch high-resolution adaptive optics observations of Arches, and derived an upper limit on the 2d velocity dispersion in agreement with the radial velocity measurements. They also discuss that astrometric follow-up observation should yield the true velocity dispersion of Arches. 
Ongoing multi-epoch astrometric monitoring of Milky Way starburst clusters will thus provide considerably improved constraints on the internal velocity dispersion, which in turn will be valuable for comparison with theoretical models.

The next generation of high-precision astrometric instruments like GRAVITY for the ESO VLTI should reveal the kinematics of stars in the very cores of the starburst clusters, provide dynamical mass estimates for the most massive stars, and possibly even trace intermediate mass blackholes hidden in the very centers of these clusters (Gillessen et al.\ 2006).

\acknowledgements 
Many thanks to Boyke Rochau and Andrea Stolte for comments and help with the figures. This paper is based in part on data relase 3+ of the UKIDSS Galactic Plane Survey (Warren et al.\ 2008, in prep). The UKIDSS project is defined in Lawrence et al (2007). UKIDSS uses the UKIRT Wide Field Camera (WFCAM; Casali et al, 2007) and a photometric system described in Hewett et al (2006). The pipeline processing and science archive are described in Irwin et al (2008, in prep) and Hambly et al (2008). 
%

\end{document}